# Enhancing Malware Detection by Integrating Machine Learning with Cuckoo Sandbox

Amaal F. Alshmarni and Mohammed A. Alliheedi
Department of Computer Science, Al-Baha University, Al Bahah, Saudi Arabia
amaalalshmarni@hotmail.com, malliheedi@bu.edu.sa

*Abstract*— In the modern era, malware is experiencing a significant increase in both its variety and quantity, aligning with the widespread adoption of the digital world. This surge in malware has emerged as a critical challenge in the realm of cybersecurity, prompting numerous research endeavors and contributions to address the issue. Machine learning algorithms have been leveraged for malware detection due to their ability to uncover concealed patterns within vast datasets. However, deep learning algorithms, characterized by their multi-layered structure, surpass the limitations of traditional machine learning approaches. By employing deep learning techniques such as CNN (Convolutional Neural Network) and RNN (Recurrent Neural Network), this study aims to classify and identify malware extracted from a dataset containing API call sequences. The performance of these algorithms is compared with that of conventional machine learning methods, including SVM (Support Vector Machine), RF (Random Forest), KNN (K-Nearest Neighbors), XGB (Extreme Gradient Boosting), and GBC (Gradient Boosting Classifier), all using the same dataset. The outcomes of this research demonstrate that both deep learning and machine learning algorithms achieve remarkably high levels of accuracy, reaching up to 99% in certain cases.

## I. INTRODUCTION

In the digital age, the ceaseless evolution of malware remains an omnipresent threat to individuals, organizations, and society at large. Malicious software, or malware, has evolved to become highly sophisticated, elusive, and continually adapted to evade traditional detection mechanisms. Amid this relentless onslaught, the fusion of deep learning techniques with the dynamic analysis capabilities of Cuckoo Sandbox emerges as a beacon of hope in the field of cybersecurity. This paper embarks on a transformative journey by introducing a groundbreaking malware dataset, meticulously curated through dynamic analysis using Cuckoo Sandbox, to drive innovation in malware detection.

Deep learning has emerged as a powerful force in various domains, including computer vision, natural language processing, and speech recognition. Its application to malware detection is compelling, as it enables the automatic extraction of intricate features and behavioral patterns exhibited by malware. However, the performance of deep learning models is inexorably tied to the quality and diversity of the data on which they are trained. Conventional malware datasets often fall short in providing the breadth and depth required to effectively combat emerging malware threats.

To bridge this gap, this paper pioneers a methodology that leverages the dynamic analysis capabilities of Cuckoo Sandbox, a widely adopted and versatile malware analysis tool. Cuckoo Sandbox simulates the execution of suspicious files within a controlled environment [1]. Observing their behavior and interactions with the system. This dynamic approach offers an unparalleled opportunity to capture the nuanced tactics and evasion strategies employed by malware, making it an ideal partner for deep learning-based malware detection. The focal point of this paper revolves around the creation of a comprehensive and timely malware dataset, meticulously constructed through the detailed analysis of malware samples using Cuckoo Sandbox.

Since 1988, computer security breaches have increased dramatically. All malicious software that infiltrates a computer system without the user's knowledge is referred to as" malware." The terms" malicious software" and" software" were combined to create this term. Malware is a significant concern in today's technological environment because it continues to increase in size and complexity. The proliferation of malware-infected websites is on the rise, posing a significant challenge to organizations that attempt to mitigate the problem. The situation is becoming increasingly concerning and can escalate beyond manageable levels, most malware infects computers when downloading data from the Internet [2]. Statistics show that many malwares and potentially unwanted applications (PUA) have been identified in recent years [3].

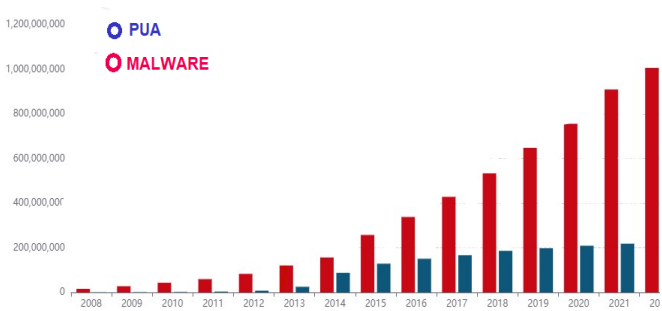

**Fig. 1**. Number of Malwares Recent Year [3]

The focus on creating methods for detecting malware and its components has increased considerably over the past few decades. Malware detection refers to the steps taken to identify malicious software. Malware can be found using one of these methods: signature-based, anomaly-based [4].

Signature-based malware detection supports a signature database and identifies malicious software by comparing suspicious patterns with those in database. Although this method can accurately detect known malware, it does so at the expense of some system resources. This method is difficult because it is useless against zero-day attacks [4]. When signature-based detection has not yet been able to identify previously undiscovered or innovative malware, anomaly-based detection can be especially helpful. However, the high false-positive rate is a significant obstacle to anomaly- based identification. If a pattern of behavior deviates too much from the norm, the system may flag it as suspicious [4]. Assurance of a computer system requires the identification of potential malware threats. However, detection of malware alone is not enough; one must also grasp its operations. Malware analysis involves uncovering the source, functionality, and repercussions of malware on a computer system. Security analysts rely on malware analysis, as it can uncover vulnerabilities within a system. There are several techniques for malware analysis, but they often fall into one of two categories: static, dynamic [5].

Static analysis is one of the earliest and most widely used methods to collect information on suspected malicious software without executing it [6]. Dynamic analysis, in contrast to static analysis, involves running suspected malware in a controlled environment and seeing how it behaves [7]. One of the methods used to detect malware is deep learning the goal of deep learning is to find hidden patterns and characteristics in data, such as photos, sounds, and texts, which involves learning abstract representations of data or observations through network layers [8]. David and Netanyahu [9] presented a study on automatic malware signature generation and classification using deep learning, called Deep Sign. To produce signatures and categorize malware, the study used a deep neural network. Strengths of the study include the use of deep learning for signature generation, which can improve detection rates, and the ability to classify multiple malware families. However, the study had a limited dataset, which may affect the generalizability of the results to real-world scenarios. Deep Sign has been tested on 1800 malware samples without benign applications and has achieved a 96.4% accuracy rate. The supervised ML model is provided by Pajouh et al. in [10]. This model estimated the frequency of each library call to find malware on Mac OS X using a kernel-based support vector machine (SVM) and weighting techniques. It was discovered that the DR was 91%. The greatest strength of the study is the novel feature selection method it developed and confirmed in several data sets. However, it is restricted to Mac OS X and cannot shed any light on malware detection on other platforms. To detect malware in API calls, Liu et al. [11] created a deep learning method. The API call chain of malicious programs was retrieved using the Cuckoo sandbox. The original sequences of API requests were recovered by sorting and filtering the duplicates. Large data sets with 21,378 samples were used to evaluate BLSTM against GRU, BGRU, LSTM and Simple RNN. According to the results of the tests, BLSTM is the most effective malware detector, with an accuracy of 97.85%.

An API call-tracking dynamic signature-based malware detection method is proposed by Savenko et al. [12] The suggested method uses dynamic signature extraction from API call sequences to detect malicious actions. The method was tested in 3,235 samples, showing a detection rate of 97. 32 %. The dynamic analysis used in this method makes it so effective in finding new forms of malware. Ijaz et al. [13] analyzed common ML algorithms used in malware detection. The inquiry process can be classified as static or dynamic, depending on its nature. According to the findings, static malware analysis has a 99. 36% success rate, while dynamic malware analysis only achieves a 94.64% success rate. Together, static and dynamic analyzes significantly improved malware detection accuracy in this investigation. The work presents the following contributions:

- The creation of a novel behavioral dataset comprising API call sequences extracted from both benign and malware files.
- An evaluation of the performance of both machine learning and deep learning algorithms to demonstrate their effectiveness in malware detection.
- Achieving a high level of accuracy in all algorithms, surpassing 90%.

## II. METHODOLOGY

This section will begin by describing the data collection and preprocessing steps, including the sources of malware and benign software samples used in the study and the methods used to extract relevant features from these samples.

We collected 2576 malware [14] and 1080 goodware samples [15] These samples were then analyzed using Cuckoo Sandbox, which executed the malware in a controlled environment and recorded all the activities performed by the malware. During the analysis, Cuckoo Sandbox generated detailed reports that included information such as file type, size, network traffic, system activity, and behavior. We exported these reports and used them to create a new malware dataset. Relevant features were extracted from the reports using JavaScript, and the resulting data set was converted to a suitable format for machine learning algorithms. All 3656 previously extracted reports are processed at this stage. Before the raw characteristics of the behavior reports can be utilized to train and test the classification algorithm, they must be categorized and cleaned up. A local directory on the host system is used to store all completed JSON behavior reports. A normal JSON report has many pieces of information, and the" Behaviors" object is where all the" behavioral" characteristics are kept. We just care about extracting API call sequences from files, good or bad. Since JSON preserves the raw characteristics of API requests, it is incompatible with deep learning techniques. Therefore, they must be processed to get a numeric representation of each file's API call attributes that can be understood by deep learning models. The process involves two main stages. The first stage deals with processing the JSON file to generate a numerical representation of each API call feature. The second stage involves processing all JSON reports to generate a comma separate value (CSV) file. Figure 2 presents various steps that were performed to accomplish the processing stage.

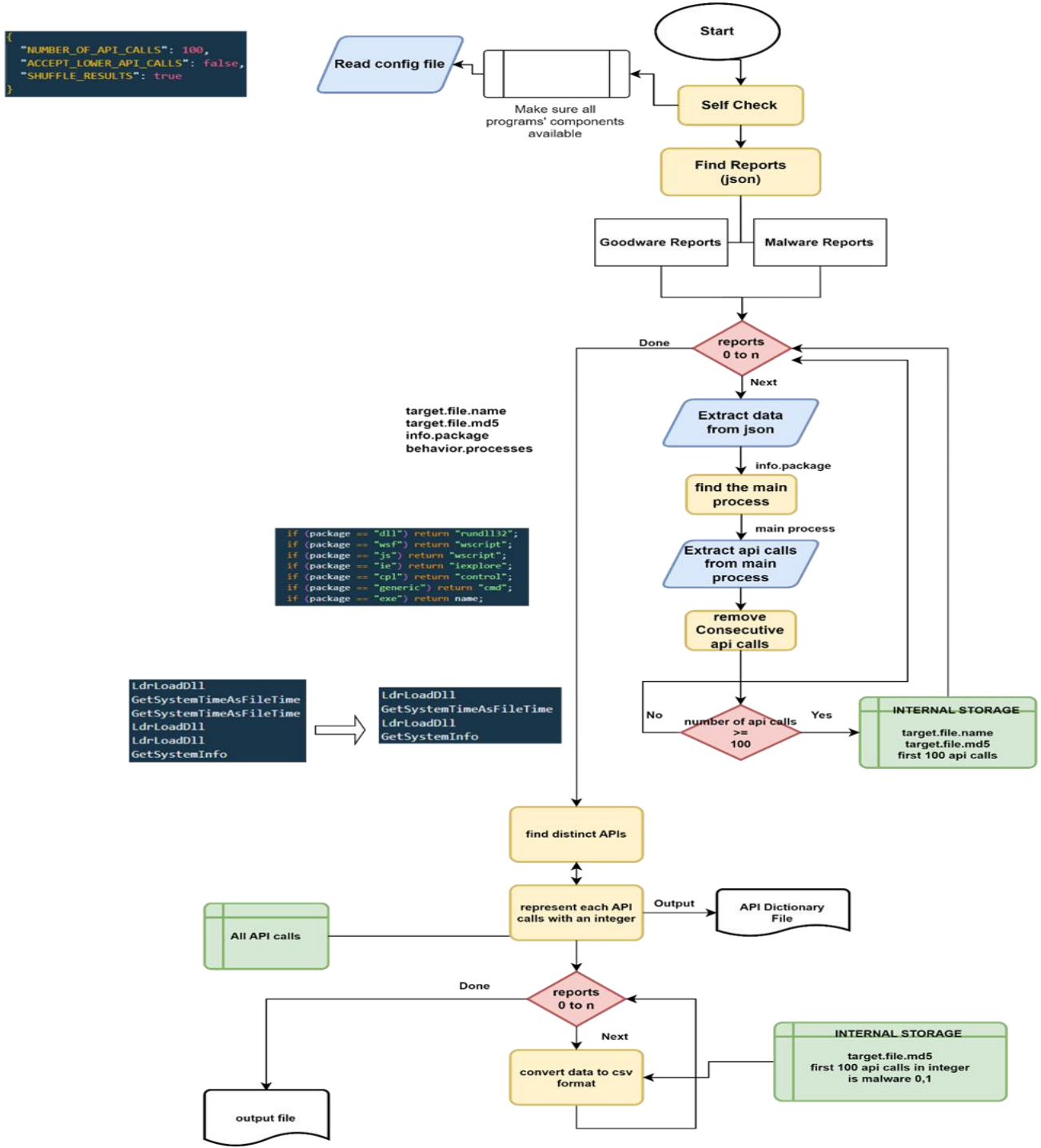

**Fig. 2**. Pre-processing JSON behavioral reports

After extracting the data, we processed this data to ensure that it did not contain missing values, in addition to solving the problem of unbalanced data. We have used deep learning algorithms such as CNN (convolutional neural network) and RNN (recurrent neural network) in this research. In addition, we compared the efficacy of deep learning algorithms with traditional machine learning algorithms.

### A. Deep Learning Algorithms

In this section, we will talk extensively about the deep learning models used in this to research detect malware. CNN is a deep neural network that is used to study the visual world to classify images, recognize images, and find objects. CNN will test and train the data that will go through convolutional layers with filters, pooling layers, and fully linked layers. It is a network where every neuron in every layer is linked to every other neuron in every other layer. CNN can be used to tell the difference between the images that are given as input and to give weights and biases to certain parts of the picture or to many other things. CNN can figure out the traits and filters. But it can also be applied to other types of data, such as malware binaries [16]. As for recurrent neural networks (RNNs), the output of a previous time step is used as input for the current time step. Recurrent neural networks (RNNs) have a memory function that retains all input data that has been provided to them. The hidden state is a crucial attribute that retains sequential information, and multiple hidden layers are available. Upon receiving input, the initial hidden layer will be activated. Subsequently, these activations will be transmitted to the subsequent layer, and the resulting activations will be relayed to yet another layer to produce an output. The process of assigning categories to each concealed layer is accomplished through the utilization of biases and weights. The network under consideration exhibits a linear graph structure among its nodes and belongs to the class of artificial neural networks characterized by a temporary sequence. The temporal exhibit's dynamic behavior is facilitated by the artificial neural network category. The internal memory of recurrent neural networks (RNNs) is utilized for the processing of sequence inputs. The implementation of a feedback loop in a recurrent neural network enables it to perceive and retain information that is connected to the previous time step [17].

### B. Machine Learning Algorithms

- Random Forest (RF): The machine-learning method RF is widely used for a variety of tasks, including classification, regression, and others. A different selection of training data and characteristics is used to build each decision tree. Taking the average opinion of all the trees in the forest yields the final forecast [18].

- Support Vector Machine: SVM are based on the notion of a margin, which is the hyperplane that separates two data classes [19].

- K-Nearest Neighbors (KNN): The fundamental concept underlying the KNN algorithm is to predict the class of a new data point using the classes of its adjacent neighbors in the feature space. K in the KNN algorithm represents the number of neighbors to consider [20].

- Gradient Boosting (GB): It is an ensemble technique that integrates several weak models into one robust model [21].

- Extreme Gradient Boosting (XGBoost): XGBoost is an extension of the conventional gradient boosting algorithm that combines gradient descent with decision tree learning. It employs a technique known as" boosting "to produce a robust model by repeatedly incorporating feeble models into the ensemble. Each weak model is trained on the residuals of the preceding model, allowing the ensemble to focus on instances that are difficult to predict [22].

The grid search across the given hyperparameter space was performed using GridSearchCV. Dictionary parameters include the validated grid search hyperparameters. The prediction technique is used to create predictions for the test set and the optimum estimator found during the grid search is used to fit the model to the complete training set. The classification report and ROC-AUC score are then used to assess the model's efficacy.

### III. RESULTS

The following metrics are calculated to evaluate the performance of the system or solution developed. We may compare different techniques using these metrics and determine whether the technique is superior. Accuracy is the ratio of valid predictions across all samples to the total sample size.

- The accuracy calculation formula is:

$$\frac{TP + TN}{TP + TN + FP + FN} \quad (1)$$

- Precision is defined as the proportion of predicted positive samples that are positive. The following is the precision formula:

$$\frac{TP}{TP + FP} \quad (2)$$

- The True Positive Rate is another term for recall (TPR).) The recall formula is as follows:

$$\frac{TP}{TP + FN} \quad (3)$$

- F1-Score is a unique statistic for assessing a model's performance that combines a harmonic measure of recall and precision:

$$2 \times \frac{\text{recall} \times \text{precision}}{\text{precision} + \text{recall}} \quad (4)$$

  o Area under the curve: The area under the curve (AUC), also known as the area under the ROC curve, summarizes the

performance of a binary classifier at various thresholds. Finding the value involves taking the ROC and calculating its area. The AUCs can be anywhere from 0 to 1. (If our AUC score increases it means that our classifier is more accurate) A higher AUC score indicates that our classifier is more accurate at predicting positive and negative examples [16].

A. Experiment 1

In the previous section we talked about how we created a new dataset that contained 2576 malware and 1080 goodware , after that we applied algorithms of deep learning, such as CNN and RNN then standard machine learning techniques such as SVM, RF, KNN, XGB, and GBC On the dataset we created. We repeated the following experiment on all of the aforementioned algorithms:

- Randomly select training data 80% and test data 20%.
- Optimal hyperparameters using grid search in machine learning techniques.

The results were as shown in the following figure.

**Table.1** Results of Experiment 1

| Method | Accuracy (%) | Precision (%) | Recall (%) | F1 score (%) | ROC-AUC score (%) |
|---|---|---|---|---|---|
| CNN | 98 | 98 | 98 | 98 | 97 |
| RNN | 99 | 98 | 99 | 99 | 98 |
| SVM | 91 | 91 | 91 | 91 | 96 |
| KNN | 92 | 92 | 92 | 92 | 95 |
| XGB | 93 | 92 | 93 | 92 | 97 |
| RF | 95 | 95 | 95 | 95 | 98 |
| GBC | 95 | 95 | 96 | 95 | 96 |

B. Experiment 2

To improve the machine learning algorithms' level of performance, we chose a different data set and worked on it. This data set is a part of our study on malware detection and classification using deep learning. 1,079 API call sequences for good ware and 42,797 API call sequences for malware are present. According to data derived from the 'calls' portion of Cuckoo Sandbox reports, each API call sequence consists of the first 100 consecutive, non-repeated API requests connected to the parent process [23]. We selected this data because it comprises more samples than other set ever used to assess the efficacy and accuracy of all algorithms. The following figure shows that the results of the deep learning algorithms obtained the same results compared to the data built. As for machine learning algorithms, their performance is improved by increasing the number of samples in the data set, the model has a greater chance of observing a distribution that is representative of all classes and can therefore learn to make more accurate predictions for all classes.

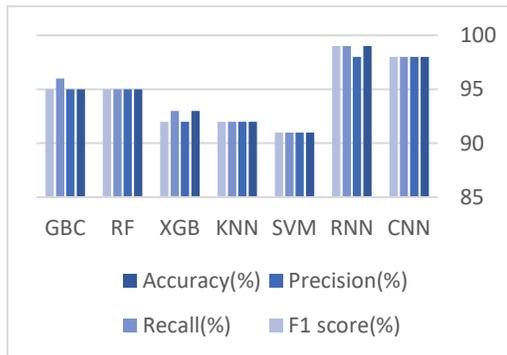

**Fig. 3** Results of Experiment 1

The effectiveness of deep learning algorithms is tested using conventional machine learning techniques including XGB, GBC, KNN, RF, and SVM. High degrees of accuracy have been attained by all the machine learning and deep learning techniques where it was the highest accuracy RNN and among them all, SVM had nearly the lowest accuracy.

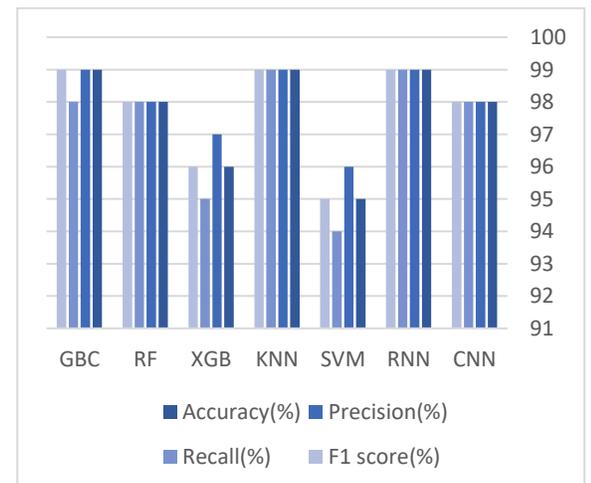

**Fig. 4** Results of Experiment 2

**Table.2** Results of Experiment 2

| Method | Accuracy (%) | Precision (%) | Recall (%) | F1 score (%) | ROC-AUC score (%) |
|---|---|---|---|---|---|
| CNN | 98 | 98 | 98 | 98 | 98 |
| RNN | 99 | 99 | 99 | 99 | 99 |
| SVM | 95 | 96 | 94 | 95 | 96 |
| KNN | 99 | 99 | 99 | 99 | 98 |
| XGB | 96 | 97 | 95 | 96 | 97 |
| RF | 98 | 98 | 98 | 98 | 98 |
| GBC | 99 | 99 | 98 | 99 | 99 |

## IV. DISCUSSION

Traditional antivirus programs lack the ability to identify and prevent rapidly emerging and complex forms of malware, which have become a significant threat to computer systems and networks. The development of deep learning has opened new possibilities for malware detection and classification utilizing the strength of neural networks. In this study, a deep learning-based malware detection system was developed and evaluated using a large dataset of malicious and benign files. The preceding section elaborated on the methodology and results of the evaluation, which obtained high detection rates and demonstrated the efficiency of the proposed system. This section will discuss the implications of the findings, highlighting the strengths and limitations of the approach. The evaluation results of the deep learning-based malware detection system show how effective it is in finding malware. The deep learning model outperformed many machine learning malware detection techniques, achieving accuracy rates of 98 percent, precision rates of 98 percent, recall rates of 98 percent, and an F1-score of 98 percent in CNN, and accuracy rates of 99 percent, precision rates of 99 percent, recall rates of 99 percent, and an F1-score of 99 percent in RNN this is because the data is linked to a specific time series. The findings imply that a variety of machine learning and deep learning methods can successfully identify malware. However, a few hyperparameters, including batch size, influence the system's effectiveness, necessitating careful tweaking for optimum performance. This outcome was produced using Adam Optimizer because it denotes optimal learning at a faster pace, the Adam optimizer is a stochastic gradient descent approach based on adaptive learning rate optimization.

## V. CONCLUSION AND FUTURE WORK

In this study, we have used a deep learning-based malware detection method and evaluated its effectiveness using a variety of malware sample datasets. As a result, we will summarize the key conclusions of our study in this section and talk about potential future research to increase the efficacy and dependability of deep learning-based malware detection. The project's main goal is to use extracted API call sequences from the dataset we created and use deep learning and machine learning algorithms to detect malware. We have used various machine learning algorithms including SVM, KNN, RF, GBC, and XGB. Although these algorithms are not well known for their effectiveness on such data, we have achieved high degrees of accuracy using these algorithms. In addition, to verify the effectiveness of deep learning techniques, we have compared and evaluated the performance of CNN and RNN, two deep learning algorithms. Although both algorithms have shown high accuracies, we found that RNN outperformed CNN by reaching an accuracy of 99%. Future work will focus on growing the data set to include more varieties of malware to be used in multiple classification problems.